# The Structure of NGC 1976 in the Radio Range

T. L. Wilson

*US Naval Research Laboratory, Code 7210, Washington, DC 20375*

*tom.wilson@nrl.navy.mil*

S. Casassus

*Departmento de Astronomia, Universidad de Chile, Casilla 36-D, Santiago, Chile*

Katie M. Chynoweth

*National Research Council Postdoctoral Research Associate,
US Naval Research Laboratory, Code 7213,
Washington, DC 20375*

**ABSTRACT**

High angular resolution radio continuum images of NGC 1976 (M42, Orion A) at ν=330 MHz (λ=91 cm), 1.5 GHz (20 cm) and 10.6 GHz (2.8 cm), have been aligned, placed on a common grid, smoothed to common resolutions of 80" (=0.16 pc at 420 pc) and 90" (=0.18 pc) and compared on a position-by-position basis. The results are <u>not</u> consistent with a single value of $T_e$. Rather, there is a significant variation, from $T_e$ =6000K in the low intensity, extended region to $T_e$=8500K in the higher intensity, more compact region. The best fit to the data is a multi-layer model obtained from radio recombination line (RRL) data. An estimate of temperature fluctuations from the model yields $t^2$=0.003. This is a factor of 10 lower than fluctuation values from optical $O^{++}$ line data.

*Subject headings:* ISM: individual (NGC 1976, M42, Orion A) - ISM: HII regions- ISM: early-type-stars- ISM: element abundances

1. Introduction

The HII region NGC 1976 (or M42, Orion A) is the most studied such source in our galaxy. At a distance of 420 pc (the average obtained from Sandstrom et al. 2007, Menten et al. 2007 and Hirota et al. 2007), this is nearest site of high mass star formation. The HII region itself has been studied extensively in the infrared, radio, and optical ranges (see, e.g. O'Dell 2001, 2003). Behind and in contact with NGC 1976 is the much more massive Orion molecular cloud. The molecular cloud remains stationary while NGC 1976 is a very thin region expanding toward the Sun (see, e.g. O'Dell 2003) with a boundary traced by molecular fragments such as CN (Rodriguez-Franco et al. 2001). The HII region has been modeled using radio continuum and radio recombination line (hereafter RRL) data. The observable parameters of such models are the electron densities, $N_e$, electron temperatures, $T_e$, radial velocities V, turbulent linewidths, $\Delta V_{turb}$, and ionization structure, averaged along the line-of-sight (LOS) and over the telescope beam. For models of NGC 1976, the geometry is a set of face-on cylindrical layers (Lockman & Brown 1975, hereafter LB; Shaver 1980). Then the LOS averages can be converted into a three dimensional distribution, to the limit set by the angular resolution. These models can be compared with data from infrared or optical wavelengths (see, e.g. Afflerbach et al. 1996, Afflerbach, Churchwell & Werner 1997). A full understanding of radio $T_e$ determinations for a sample of nearby regions would aid interpretations for more distant sources.

## 2. Models based on RRL's

The first internally consistent RRL model of an HII region was by Brocklehurst & Seaton (1972, hereafter BS). For the analysis of RRL emission, detailed radiative transfer is needed. This was first carried out by BS for a spherical geometry. Later LB, Shaver (1980) and Wilson & Jaeger (1987, hereafter WJ) refined the model by using a multi-layer, slab geometry. For NGC 1976, a geometry where diameters are much larger than LOS depths is needed since the RRL data requires local densities in excess of the RMS average densities (see Appendix). Most of the RRL data were taken toward the continuum maximum with a variety of angular resolutions. Decisive tests of these models require determinations of $T_e$ from RRL maps. In the face-on slab geometry, the outer parts of the nebula are closer to the Sun. The model of LB predicted a rise in $T_e$ in the outer parts of NGC 1976, while the Shaver (1980) model is isothermal. The data taken by WJ showed that the value of $T_e$ is lower further from the center of NGC 1976. These results led to the WJ model in which the most extended, lower density layer is closer to the observer, with $N_e$ and $T_e$ increasing with increasing distance from the observer. The maximum value of $T_e$ reached 8500 K for the compact dense gas at the interface to the background molecular cloud to the rear. The WJ model is consistent with the RRL data taken from 600 MHz to 100 GHz toward the source peak, and 22 GHz radio continuum scans in Right Ascension and Declination, and maps in RRL's. Thus this model is the starting point for an analysis of continuum images at frequencies of 330 MHz (Subrahmanyan, Goss & Malin 2001; resolution 79" by 65") 1.5 GHz (Felli et al. 1993; resolution 28") and 10.6 GHz (Subrahmanyan et al. 2001; resolution 90").

The WJ model is based on the radiative transfer method of BS (in a radiative transfer computer code from C. M. Walmsley) and incorporates the face-on layered geometry introduced by LB and improved by Shaver (1980). A major change from earlier models was the introduction of a decrease in $T_e$ from back to front of the source. This was based on the measurement of $T_e$ at four positions 4' from the peak in the hydrogen 110α line (ν=4.87 GHz ) and H139β (ν=4.79 GHz). At these positions, non-LTE effects are expected to be negligible since the continuum intensities at 4.8 GHz are small, so τ is <0.05 (see Eq. A2 & A3). From these data, the derived $T_e$ values away from the continuum peak are smaller than toward the peak. In the layer geometry, this is interpreted as a decline in $T_e$ in the outermost layer. This fall-off in $T_e$ was an important difference between the models of WJ, LB and Shaver (1980). The WJ model was originally a 5 layer face-on slab geometry. To better fit continuum images, the four innermost layers were divided into two parts, each with one-half the LOS depth, the same $N_e$ and $T_e$, and the layer nearer the sun having a larger extent perpendicular to the LOS. The outermost layer, closest to the Sun, was left single. Thus the five layer model grew to nine layers. As shown by WJ, the RRL data toward the peak could be fit using the LB, Shaver (1980) or WJ models, so only results from RRL mapping data allowed a decision between these.

### 3. Radio Continuum Data

In principle, $T_e$ can be determined from radio continuum data. The relation for a single isothermal region in the Rayleigh-Jeans regime is (see e.g. Wilson et al. 2009):

$$T_B = T_e \left(1 - e^{-\tau_{ff}}\right) \quad (1)$$

where $\tau_{ff}$ is the free-free optical depth (see Appendix for a discussion of $\tau_{ff}$). $T_B$ is the true brightness temperature. This will be the measured value if the source is much larger than the telescope beam.

A number of methods have been used to estimate values of $T_e$. The first is to solve Eq. (1) for the case when $\tau \gg 1$. For NGC 1976, this requires measurements at a low frequency. For a discrete source, the value of $T_B$ is not directly measured, but this can be obtained from measurements of the peak intensity as main beam brightness temperature, $T_{MB}$, accurate calibrations and measurements of the beam and source size. The usual assumption is that the beam and source have Gaussian shapes. From single telescope data, the core of NGC 1976 has a rather large intensity and an angular size of ~2.5', but there is lower intensity emission from the nebula over a total extent of ~30' (see, e.g. Wilson et al. 1997, Dicker et al. 2009). From the variation of flux density with frequency, the value of $\tau_{ff}$ of the core of NGC 1976 is larger than unity below 1 GHz, so from Eq. 1, high angular resolution measurements for frequencies at or below 1.5 GHz can provide a good determination of $T_e$. Mills & Shaver (1968) reported a value of $T_e = 7600 \pm 800$ K using a 3.2' beam at 408 MHz. Subrahmanyan et al. (2001) measured a value of 3.84 Jy with a 65" by 79" beam at 330 MHz using the Very Large Array (VLA). This yields a peak temperature of $T_{MB} = 8400$ K. We estimate that the error from noise is less than 30K. Felli et al. (1993) measured a peak intensity of 6.3 Jy with a 28" beam at 1.5 GHz, using the VLA in the C and D configurations. This yields a peak temperature of $T_{MB} = 4600$ K. We estimate that the error from noise is less than 12K. However at 1.5 GHz, the value of $\tau_{ff}$ is perhaps 20% uncertain, so the value of $T_e$ has at least this uncertainty (Eq. 1).

A second method consists of determining $T_e$ and $\tau_{ff}$ from a fit to the source flux density as a function of frequency. This fit is based on an application of Eq. 1, Eq. A5 and A6. In addition, there must be a number of measurements over a wide range of frequencies (see, e.g. Terzian & Parrish 1970). The errors in such determinations are often underestimated since the zero levels, limits of integration to determine the total flux density and calibrations may be inaccurately determined. A smaller effect is that the flux density of the nearby source NGC 1977 must be separated from the total integrated flux density.

A third method was presented by Dicker et al. (2009). These authors introduced a more elaborate approach. In this, $T_e$ is determined using resolved, calibrated images taken at two frequencies. These are smoothed to a common resolution, aligned and placed on a common grid. Then, intensities are plotted on a position-by-position basis to determine unique values of $T_e$. The analysis does not depend on geometry or a knowledge of $\tau_{ff}$. In Fig. 1, we follow the scheme of Dicker et al. (2009), plotting the intensities for 330 MHz (data of Subrahmanyan et al. 2001) and 1.5 GHz (data of Felli et al. 1993). In our approach, we did not allow for a DC offset or intensity calibration error (as was done by Dicker et al. 2009). The data for both wavelengths were smoothed to 80", the maxima aligned and placed on a common grid. In addition, there are plots for isothermal models for $T_e$=4000 K, 6000 K and 9000 K. From these results, there is a systematic difference with respect to the models of constant $T_e$, so these data are *not* consistent with a single value of $T_e$. For lower intensities, the data agree with a $T_e$ value of ~6000 K, while for higher intensities, the agreement is with a $T_e$ value of ~8500 K. Given the averaging along the line of sight, it is difficult to deduce a model from this data, but one *can* test models on the basis of agreement with the data plotted in Fig. 1.

The best choice is a comparison with the WJ model, since this agrees with RRL mapping data. From a by-eye comparison, the WJ model is more consistent than plots for a constant $T_e$. The agreement between model and data was improved by lowering the $T_e$ values in the three outermost layers by 8%. In Table 1 the parameters of the improved version of the WJ model are given; the distance to the source is assumed to be 420 pc. A side view of this model is given in Fig. 2 This improved model is compared with data in Fig. 3, where we show this model as a dash-dotted line. In Fig. 4 we plot intensity versus intensity for the 1.5 GHz and 10.6 GHz data. From this plot, the difference between the improved WJ and isothermal models is rather minor. The value of $T_e$=11376 $\pm$1050 K reported by Dicker et al. (2009) was based on data taken at 1.5 GHz and 21.5 GHz, but from Fig. 4, data taken at frequencies larger than 1.5 GHz do not provide strong constraints on $T_e$ values. In addition, the maximum intensities reported by Dicker et al. (2009) exceed those of Felli et al. (1993; 1.5 GHz) and Wilson & Pauls (1984; 23 GHz) by a factor of 1.3; we believe that this factor reflects an inconsistency in the value of the beam areas used to obtain the intensity units. For 1.5 GHz, there is excellent agreement between the integrated flux densities reported by Felli et al. (1993) and van der Werf & Goss (1989). However, the maximum value of integrated flux density at 1.5 GHz from the improved WJ model is 28% larger than these flux densities, so some more extended structure has not been recorded by the VLA. If the missing flux density is uniformly distributed over 30', the intensity is 22 mJy beam$^{-1}$ or 0.2 MJy ster$^{-1}$, for a 28" resolution. For the model in Table 1, this additional intensity will have a negligible effect.

The shapes of the curves in Figs. 1 and 3 are quite different from those of our Fig. 4 or Fig. 7 of Dicker et al (2009). This is because the optical depths in NGC 1976 are significant at 330

MHz. In the improved WJ model, the sum of the optical depths, $\tau$, of layers 7 and 8 is 1.9, while at frequencies higher than 1.5 GHz, the value of $\tau$ for these layers is <0.1. The inclusion of the 330 MHz data is a crucial input for any test of the models of NGC 1976.

## 4. Discussion

Early measurements of RRL's gave values of $T_e$ that were markedly lower than values determined from collisionally excited lines of $O^{++}$ in the optical (see, e.g. Goldberg 1966). In one response to these results, Peimbert (1967) pointed out that the collisionally excited lines (CEL's) are biased toward high values of $T_e$ while the radio recombination lines are biased toward lower values. To compare these radio and optical data, Peimbert (1967) introduced an unbiased temperature, $T_0$ and a factor to characterize the fluctuations, $t^2$. The value of $T_0$ is defined as

$$T_0(N_i, N_e) = \frac{\int T(r)\, N_i(r)\, N_e(r)\, d\Omega dl}{\int N_i(r)\, N_e(r)\, d\Omega dl} \qquad (2)$$

Where T is the electron temperature, $N_i$ is the ion density, and r is the distance along the line of sight. The integration is over the angular resolution. The fluctuation in $T_0$ is given by:

$$t^2 = \frac{\int (T(r) - T_0)^2\, N_i(r)\, N_e(r)\, d\Omega dl}{T_0^2 \int N_i(r)\, N_e(r)\, d\Omega dl} \qquad (3)$$

Using the improved WJ model (Table 1), there can be two extreme values of $t^2$. The first is for a very small angle, and the second for integration over the entire source. The values for the very small angle are $T_0$=8250K and $t^2$=3 10$^{-3}$. For the entire source these are $T_0$=7600K and $t^2$=10$^{-2}$. The very small angle result is most relevant for comparisons with optical results. According to Peimbert (1967), the value of $T_0$ from radio recombination lines, T(RRL), is weighted as:

$$T(RRL) = T_0 \left[1 - 1.42\, t^2\right]$$

For the optical collisionally excited lines (CEL) such as the optical [OIII] lines, the result is:

$$T(CEL) = T_0 \left[1 + \frac{1}{2}\left(\frac{90800}{T_0} - 3\right) t^2\right]$$

For our value of $t^2$, T(RRL)=T(CEL). Our values are ~10% of the values given by Esteban et al. (2004), who estimated that $t^2$ is between 0.02 and 0.03 with a favored best value of 0.022 ±0.002. This cannot be reconciled with our model, but indicates that structure on a finer scale than our 80" to 90" resolution is present. Higher values of $T_e$ could arise in higher density clumps. Tsamis et al. (2008) had discussed effects that lead to finite $t^2$ values in Planetary Nebulae. The possibilities are fluctuations in $T_e$ or in abundances of N or O. For NGC 1976, the first is more likely.

Baldwin et al. (1991) report a value of $T_e$ =12000 K from measurements of singly ionized Sulfur, [S II]. These arise from the ionization front of the HII region; if so, these are in conflict with our model. The RRL data for the densest region give lower values, with $T_e$ ranging from 8300K to 8600K; these data were taken with ~1' beams, so are averages over a larger region. O'Dell & Harris (2010) find a maximum value of $T_e$=8610 ±640 K, and review previous optical results.

Other approaches to the question of element abundances and the value of $T_e$ are possible. For example, Rubin et al. (1998) have used the Hubble Space Telescope (HST) to measure ultraviolet lines of N II] and [O II]. By forcing agreement between abundance ratios from optical and UV measurements, they find $T_e$=9500 K, $t^2$=0.032 and $N_e$=7 $\cdot$ $10^3$cm$^{-3}$. From the values of $N_e$, these lines must be associated with the model layers 1 to 4. However, this approach depends on combining measurements in the optical and UV, so on many factors. This may contribute to values differing from our determination of $T_e$. Except for the data of Baldwin et al. (1991), the data favor $T_e$ values lower than $10^4$ K throughout NGC 1976. The model in Table 1 must be considered the best model, since this predicts a fall-off of $T_e$ in the foreground region, layer 9, with the largest $T_e$ values in the rear part of the H II region.

Other than NGC 1976, only a few sources have been studied in infrared fine structure lines (see Afflerbach et al. 1997) but the source sample will be enlarged by Herschel Space Observatory and the start of science flights with the Stratospheric Observatory for Far IR Astronomy (SOFIA). As a preparation for future surveys, detailed models of nearby sources are of value, especially from comparisons of different methods that allow the assessment of systematic effects.

## 5. Conclusions

The analysis of the intensities of pairs of high resolution continuum images gives strong support to a small modification of WJ model, which is based on RRL data. The WJ model is based on data averaged over 40" (=0.08 pc at 420 pc), so represents large scale trends. In this model, there is a significant variation, from $T_e$ =6000K to $T_e$=8500K. The lowest $T_e$ arises from the most extended, lowest density foreground gas. The highest $T_e$ value arises from the densest, most compact region that is further from the sun and abuts the molecular cloud; this is the location of the ionization front, so $T_e$ and $N_e$ monotonically decreases from the ionization front to the region of low density gas. The model gives a $T_e$ fluctuation value, $t^2$, which is~10% of the values deduced from optical data. This is an indication that small scale structure in $T_e$ is averaged out by our lower angular resolution.

Acknowledgements: We thank W. M. Goss, R. Subrahmanyan and G. Taylor for providing the digital form of the images used in the analysis. This research was performed while an author held a National Research Council Research Associateship Award at the US Naval Research Laboratory.

## 6. Appendix

After the discovery of RRL's, it was assumed that these were emitted in Local Thermodynamic Equilibrium, LTE. If so, the integrated line-to-continuum ratio allowed a direct estimate of the electron temperature, $T_e$. Under the assumption of Local Thermodynamic Equilibrium, the expression is

$$T_e^* = \frac{1.91 \cdot 10^4}{R <g_{ff}(T_e^*)>} \left(\frac{N+}{N_e}\right) \qquad (A1)$$

where $T_e^*$ is the electron temperature in Kelvins, calculated under the assumption of LTE, R is the integrated line-to-continuum ratio in units of km s$^{-1}$, N$^+$ represents the number of ions, in this case H$^+$, while N$_e$ represents the total number of electrons from all ions, $<g_{ff}(T_e^*)>$ is the Gaunt factor (Eq. A5) and ν is the line frequency in GHz. This expression must be solved iteratively.

However, the formation of RRL's is a non-LTE process (see Goldberg 1966, Gordon & Sorochenko 2009). In the following we restrict the discussion to hydrogen RRL emission in a source such as NGC 1976. The level populations are determined by the capture of free electrons into levels with high principal quantum numbers, n, followed by radiative decay to levels n=1 or 2, followed by reionization. For high n levels, collisions play a larger role, while for low n levels, radiative decay is more important. The level populations for level n are described by the departure coefficients or *b* factors (ratios of the actual to the LTE population for a principal quantum number n). The value of *b* is a function of n, N$_e$ and T$_e$. As n approaches 1, *b* approaches zero monotonically. For the hydrogen atom, *b* is always less than unity. The *b* factor accounts for a lowering of energy level populations caused by radiative decay; for large n, *b* approaches unity. Another factor, β, is used to describe the difference in population between energy levels. The population of a higher level is larger than level n, so since kT$_e$ >> hν. This population difference leads to an amplification of the combined background consisting of continuum and spectral line emission. The amplification factor, β, depends on geometry, N$_e$ and T$_e$; *b* causes a reduction in line intensity while β causes an increase in line intensity. For n>100, collisions shift line intensity to the line wings. For NGC 1976, from the models of Shaver (1980) and WJ, at frequencies between 4 GHz (n~120) and 30 GHz (n~60) the effects of lower population and line amplification are rather small and largely compensate.

Non-LTE effects influence the intensity of RRL's. RRL emission is affected by: (1) a lowering of the energy level population and thus the line intensity by the factor $b<1$, (2) line masering which increases the line intensity, as expressed by the factor $\beta$, and (3) collisional broadening. In the formulation of BS, the relation between $T_e$ and $T_e^*$ is:

$$T_e = b\, T_e^* \left(1 + \frac{1}{2}\beta \cdot \tau\right) \cdot \frac{<g_{ff}(T_e^*)>}{<g_{ff}(T_e)>} \qquad (A2)$$

Where $\tau$ is the optical depth in both line and continuum. Accurate values of $b$ and $\beta$ are needed to determine source parameters; see BS for sample values and applications to NGC 1976. Qualitatively, effect (2) is important if a low value of $N_e$ is combined with large continuum intensity. Then this leads to large amounts of RRL masering, which is not found. Thus the local density must be raised. In the case of NGC 1976, this was done by assuming that the emission arises from thin face-on slabs. This is consistent with the HII region located at the front side of a massive molecular cloud. Effect (3) is important for frequencies $\nu < 2$ GHz in NGC 1976. Then lines are broadened by collisions with free electrons, since these atoms have larger sizes. The larger linewidths may be inadvertently removed in baseline fits, so that some of the line emission is not recorded. This effect can be reduced by plotting the line-to-continuum ratios and linewidths separately.

For continuum measurements, the simplest expression is for a uniform temperature slab, as in Eq. (1), where detailed physics is contained in $\tau_{ff}$. This is given by the classical expression:

$$\tau_\nu = 3.014 \times 10^{-2} \left(\frac{T_e}{K}\right)^{-3/2} \left(\frac{\nu}{\text{GHz}}\right)^{-2} \left(\frac{\text{EM}}{\text{pc}\,\text{cm}^{-6}}\right) \langle g_{ff} \rangle \qquad (A3)$$

here EM, the emission measure of an ionized gas. If one neglects singly ionized helium, this becomes $N_i = N_e$:

$$EM = \int_0^s N_e^2 \, ds \qquad (A4)$$

For a LOS depth "s" equal to the observed diameter, the density will be the RMS value. For the NGC 1976 geometry, the RRL data were used to determine $N_e$ and from this, the LOS depth must be much smaller than the diameter of a layer. And the classical expression for the Gaunt factor is that given by (Oster 1961):

$$\langle g_{ff} \rangle = \ln\left[4.955 \times 10^{-2} \left(\frac{\nu}{\text{GHz}}\right)^{-1}\right] + 1.5 \ln\left(\frac{T_e}{K}\right) \qquad (A5)$$

where "< >" signifies an average over velocities.

In the literature, one finds some other expressions for the Gaunt factor but these differ only by a constant factor. For example, Beckert et al. (2000) used a Gaunt factor that is 1.73 times larger than Eq. A5. Altenhoff et al. (1960) had approximated the Gaunt factor by a power law relation in $T_e$ and $\nu$:

$$\tau_\nu = 8.235 \times 10^{-2} \left(\frac{T_e}{K}\right)^{-1.35} \left(\frac{\nu}{\text{GHz}}\right)^{-2.1} \left(\frac{EM}{\text{pc cm}^{-6}}\right) a(\nu, T) \qquad (A6)$$

With the correction $a(\nu, T)$ of ~1. This expression for $\tau$ has been used for many calculations in the radio range. This is valid for $\nu < 10$ GHz, but there are significant deviations from the classical expression for $< g_{ff} >$ at higher frequencies, even for $T_e$ values of 7000 K (see Beckert et al. (2000)). Thus, one should use the classical expression for $\tau_{ff}$, that is, the combination of A3 and A5, but *not* Eq. A6, to avoid errors for frequencies >10 GHz. The exact calculation for the Gaunt factor for $T_e$ values below $10^3$ K must be calculated using the more accurate formulae, such as those summarized in Casassus et al. (2007).

For the transfer of continuum radiation through regions of differing $T_e$, one requires a more elaborate relation. For a ray through the center of NGC 1976, the WJ model listed in Table 1, gives:

$$T_B = T_e(1) \cdot \left(1 - e^{-(\tau_1+\tau_2+\tau_3+\tau_4)}\right) \cdot e^{-(\tau_5+\tau_6+\tau_7+\tau_8+\tau_9)} + $$
$$T_e(5) \cdot \left(1 - e^{-(\tau_5+\tau_6)}\right) \cdot e^{-(\tau_7+\tau_8+\tau_9)} + $$
$$T_e(7) \cdot \left(1 - e^{-(\tau_7+\tau_8)}\right) \cdot e^{-\tau_9} + $$
$$T_e(9) \cdot \left(1 - e^{-\tau_9}\right) \quad (A7)$$

For the emission from layer 9 only, this reduces to Eq. 1. For rays offset from the center, contributions from some layers may be absent since their lateral extents are different. Applying this relation to the original WJ model, for frequencies of 330 MHz and 1.5 GHz, gives the curve in Fig. 1. The improved model, with parameters in Table 1 and a side view in Fig. 2, results in the curve shown in Fig. 3. For the frequencies 10.6 GHz and 1.5 GHz, this results in the curve shown in Fig 4.

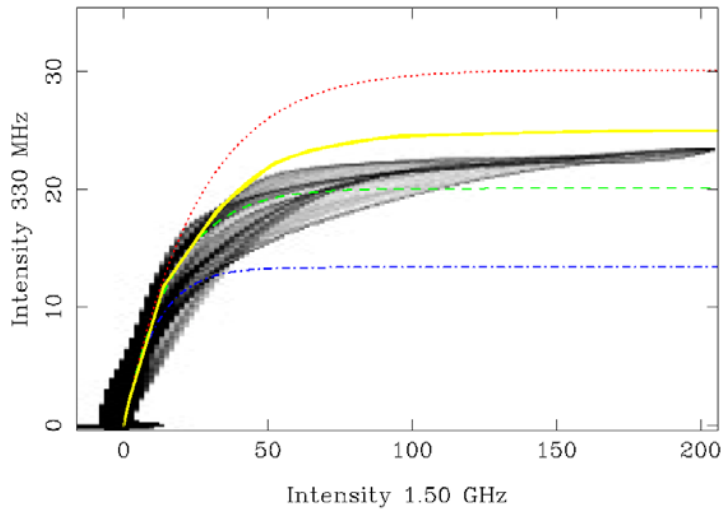

Fig.1 A plot of the relationship between radio continuum intensities for a pair of frequencies. Rather than using a scatter plot, which gives rise to crowded regions and deemphasizes outliers,

we have plotted the data as a grey-scale representation. The units are MJy per steradian. The 330 MHz data is from Subrahmanyan et al. (2001), for 1.5 GHz (Felli et al. (1993). Both data sets were smoothed to 80" resolution, with maxima aligned and data placed on the same grid points. This plot follows the method of Dicker et al. (2009). The dotted line is for a model made following Eq. 1, smoothed to 80" resolution, with a constant electron temperature of $T_e$=9000K, the dashed line is for a constant electron temperature of $T_e$=6000K, the dash-dotted line is for a constant electron temperature of $T_e$=4000K. The thicker solid line is for the WJ model (parameters in that paper).

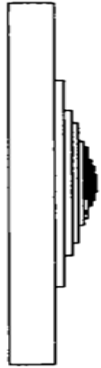

Fig. 2. A side view of the model whose parameters presented in Table 1. The observer is on the left, and layer 9 is the largest region closest to the observer. Sizes are not to scale.

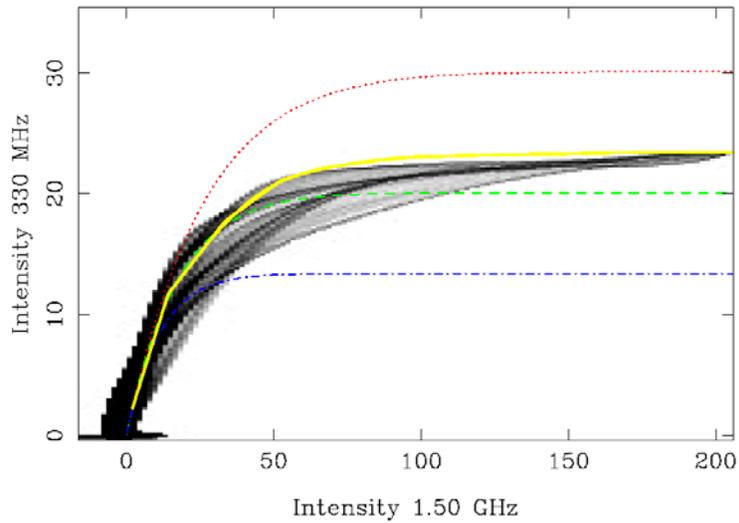

Fig.3. The data and constant $T_e$ models as in Fig. 1. The improved WJ model is shown as a thicker solid line (parameters in Table 1 and see Fig. 2).

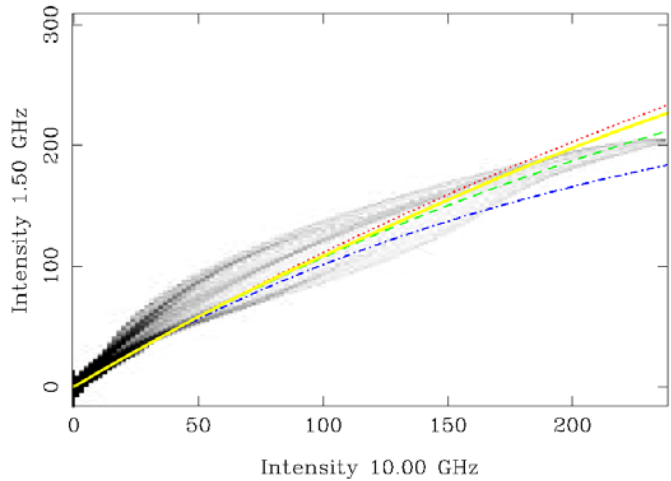

Fig 4. The data sets processed as in Fig. 1. Here the data were taken at 10.6 GHz (Subrahmanyan et al. (2001)) and 1.5 GHz (Felli et al. (1993)). Both data sets were smoothed to 90" resolution, with maxima aligned and data placed on the same grid points. The dashed, dotted and dash-dotted lines are for constant temperature models as in Fig. 1. The thicker solid line is the improved WJ model (Table 1).

**Table 1: An Improved Model of NGC 1976 from Radio Astronomy Data**

| layer | $T_e$ (K) | line-of-sight path (pc) | $N_e$ (cm$^{-3}$) | mean diameter (pc) | Fraction of He$^+$ | Turbulent Velocity (km s$^{-1}$) |
|---|---|---|---|---|---|---|
| 1 | 8500 | 6.0 (-3) | 1.2 (4) | 0.08 | 1 | 10 |
| 2 | 8500 | 7.0 (-3) | 1.0 (4) | 0.17 | 1 | 10 |
| 3 | 8500 | 9.0 (-3) | 8.9 (3) | 0.25 | 1 | 15 |
| 4 | 8500 | 1.6 (-2) | 7.5 (3) | 0.34 | 1 | 15 |
| 5 | 8000 | 2.3 (-2) | 5.4 (3) | 0.42 | 1 | 15 |
| 6 | 8000 | 2.8 (-2) | 3.7 (3) | 0.59 | 1 | 15 |
| 7 | 7000 | 3.7 (-2) | 2.7 (3) | 0.76 | 0.75 | 12 |
| 8 | 7000 | 3.9 (-2) | 1.7 (3) | 1.10 | 0.75 | 12 |
| 9 | 6000 | 2.0 (-1) | 3.0 (2) | 1.85 | 0.5 | 12 |

20